\documentclass[doublecol]{epl2} 

\title{Dynamics of helical vortices and helical-vortex rings}

\author{E. B. Sonin
}

\institute{   Racah Institute of Physics, Hebrew University of
Jerusalem, Jerusalem 91904, Israel                 
}
\pacs{67.25.dk}{Vortices and turbulence}
\pacs{47.32.cf}{Vortex reconnection and rings}
\pacs{67.30.he}{Textures and vortices}

\abstract{
The Letter considers dynamics of helical vortices and helical-vortex rings either solving directly the equations of motions of a vortex line or using canonical relations following from the Hamiltonian equations of motion. An analytical solution in elliptical integrals was found  for helical-vortex rings in the local-induction approximation. The analysis based on the canonical Hamilton relation provides a clear physical explanation for anomalous velocities of helical-vortex rings, {\em i.e.}, for suppression of the velocity and even inversion of its direction at sufficiently large amplitude of the helical distortion. The extended local-induction approximation is suggested, which provides an exact solution for the equations of motion  of helical vortices and helical-vortex rings in the limit when the small-pitch helical vortex reduces to a cylindric sheet of uniform vorticity. }

\begin{document}

\maketitle

\newcommand{\be}{\begin{equation}}
\newcommand{\ee}[1]{\label{#1}\end{equation}}
\newcommand{\bem}{\begin{eqnarray}}
\newcommand{\eem}[1]{\label{#1}\end{eqnarray}}
\newcommand{\eq}[1]{eq.~(\ref{#1})}
\newcommand{\Eq}[1]{Equation~(\ref{#1})}


\section{Introduction}

Helical vortices  were intensively studied in classical hydrodynamics \cite{Saf,Oku}. They  appear in wakes of propellers and other spinning bodies. A helical vortex is a straight vortex line with a circularly polarized Kelvin wave of arbitrary amplitude propagating along it.
One may  roll up the helical vortex into a ring keeping its helical deformation. 
Then it will be a helical-vortex ring.
An ideal vortex ring is an ubiquitous object in vortex dynamics in general and in superfluid turbulence in particular and has already been studied about two centuries. Its dynamics is well established. Recently attention was focused on dynamics of helical-vortex rings with large-amplitude Kelvin waves propagating around the ring \cite{KM,BHT,Bar2}.  This problem being interesting itself may have important implications for superfluid turbulence. An interesting outcome of studies of this object was that for Kelvin-wave amplitudes large enough the vortex ring may move in direction opposite to its moment. This phenomenon was first revealed by Kiknadze and Mamaladze \cite{KM} in numerical calculations and the perturbation theory  within the local-induction approximation. Later it was confirmed \cite{BHT,Bar2} by numerical calculations based on the Bio--Savart law.  The effect seemed to be mysterious and was called {\em anomalous vortex-ring velocity} \cite{BHT,Bra}.

This Letter addresses dynamics of helical vortices and helical-vortex rings in the local-induction approximation either solving directly the equations of motions of a vortex line or using simple canonical relations following from the Hamiltonian equations of motion. A fully analytical solution in elliptical integrals was found  for helical-vortex rings 
with periodic helical distortion of arbitrary amplitude. It confirms  a simplified approach based on the general canonical Hamilton relations. The latter approach  provides a  transparent physical explanation for the origin of the anomalous  vortex-ring velocity, which now does not deserves an adjective ``anomalous''.  Suppression of the ring velocity and the eventual inversion of its direction are related with conserved angular momentum around the ring axis, which is generated by the helical deformation of the vortex ring.  Agreement of the canonical approach with the full analytical solution (within the area of validity for the former) is a proof of its effectivity. 

The local-induction approximation assumes that a velocity of a fixed point on a vortex line is induced only by the closest segment of the line and is determined by its curvature radius. In the case of a helical vortex the closest segment is restricted by one turn of the helix. The presented analysis goes beyond the strict local-induction approximation and suggests the extended local-induction approximation, which takes into account also the velocity induced by other segments (turns) of the vortex line. Approximately estimating the effect of the other turns they are   replaced by a cylindric vortex sheet with uniform distribution of vorticity. This approach gives an exact solution of the Bio--Savart law of vortex motion in the limit of a helix of very low pitch compared to the helix radius.  Being also accurate in the opposite limit of high pitch, where the local-induction approximation is valid, the extended local-induction approximation provides a reliable interpolation  model for a broad spectrum of vortex   configurations.

\section{Helical vortex} \label{sing}

\begin{figure}
\onefigure[scale=0.35]{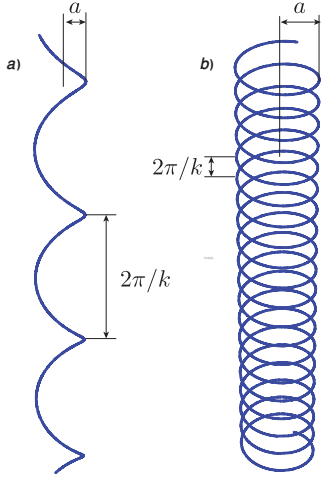}%
 \caption{Helical vortex. (a) Helix of high pitch $2\pi/k \gg a$. (b) Helix of low pitch $2\pi/k \ll a$. }  \label{f2}
 \end{figure}

Let us consider a Kelvin wave of arbitrary amplitude propagating along a straight vortex line. In the Cartesian coordinate frame with the $z$ axis coinciding with the  unperturbed vortex line, the transverse displacements in the circularly polarized Kelvin wave  are
\be
x=a \cos(kz-\omega t)\,,\qquad y=a \sin(kz-\omega t)\, .
      \ee{1}
So in the Kelvin mode the vortex line forms a helix moving along a cylinder coaxial with the $z$  axis with the pitch equal to the Kelvin wavelength $2\pi /k$ (fig.~\ref{f2}).

In the local induction approximation the equation of motion for a curved vortex line is 
\be
-\rho_s  \kappa  \hat s(\bm  R_L) \times  \frac{\upd \bm  R_L}{\upd t} = -
\frac{\mathbf  {\cal R}}{{\cal R}^2} \epsilon \, .
                   \ee{EqCur}
Here $\rho_s$ is the superfluid mass density, $\kappa =h/m$ is the circulation quantum, $\bm  R_L$ the position vector for points on the vortex line, $ \hat s(\bm  R_L)$ is the unit vector tangent to the vortex line at the point with the position vector $\bm  R_L$, $ \mathbf  {\cal R} $ is the curvature-radius vector, which is determine by the relation
\be 
 \frac{  \mathbf{ \cal R}}{{\cal R}^2}=\frac{\upd  \hat s }{ \upd l}\, ,
  \ee{2}
  $\upd l$ is the element of the vortex line length, and $\epsilon$ is the vortex line tension equal to the vortex-line energy per unit length:
  \be 
\epsilon =\frac{\rho_s \kappa^2}{ 4\pi} \ln\frac {r_m }{ r_c} =\rho_s \kappa \nu_s\, .
  \ee{3}
Here $r_c$ is the core radius,  $r_m$ is the scale to which the velocity  field $v \sim 1/r$ penetrates, and $\nu_s =\kappa \ln(r_m/r_c)/4\pi$ is the line-tension parameter.  
The curvature-radius vector at any point of the helix is directed along the radius of the cylinder while its absolute value is  ${ \cal R}= (1+k^2a^2)/k^2a$. In the cylindrical coordinate system $(z,r,\phi)$ the unit tangent vector has components $s_r=0$, $s_\phi =ak/\sqrt{1+k^2a^2}$, and $s_z =1/\sqrt{1+k^2a^2}$. Then solution of the equation (\ref{EqCur})  of motion for the vortex can be presented in two forms (auto model solutions):
\be
\phi(z) =kz-\omega t= k(z- v_z t)\, ,
   \ee{4}
or 
\be
z=\frac{\phi - \Omega_z t}{k}\, .
   \ee{5}
Thus the helix motion may be described either as  pure vertical translation with the velocity $v_z$, or pure rotation with angular velocity $\Omega_z$. According to \eq{EqCur} 
 \be 
v_z=\frac{\omega}{ k} = \frac{\nu_s k}{ \sqrt{1+k^2a^2}}\, ,
  \ee{disp}
\be
\Omega_z  =- \frac{\nu_s k^2 }{ \sqrt{1+k^2a^2}}\, .
  \ee{rot}
In the limit of small $a$ this yields  the dispersion relation $\omega=\nu_sk^2$ for the linear Kelvin wave. But the frequency decreases with increasing amplitude $a$ of the Kelvin wave.

Expressions (\ref{disp})  and (\ref{rot}) can be also derived  from the general canonic Hamilton equations 
\be
v_z =\frac{\partial E}{ \partial P_z}=\frac{\partial E/\partial a}{ \partial P_z/\partial a}\,,\qquad\Omega_z =\frac{\partial E}{ \partial M_z} =\frac{\partial E/\partial a}{ \partial  M_z/\partial a}\, . 
 \ee{can}
where $P_z$ and $M_z$ are the linear momentum and the angular momentum along the $z$ axis,  
\be
E=\frac{\rho_s \kappa^2 L}{ 4\pi} \left(\sqrt{1+k^2a^2}\ln\frac{1}{ kr_c}+\ln\frac{kR_c }{ \sqrt{1+k^2a^2}}
\right)\,,
  \ee{en}
is the energy of the vortex line, and $L$ and $R_c$ are  the hight and the radius of the container with the helical vortex. Partial derivatives in \eq{can}  took into account that the only varying parameter of the helix is its radius $a$. Indeed, imposing the periodic boundary conditions in the box of the hight $L$ the pitch $2\pi/k$ of the helix becomes a topological invariant. The first term in the expression (\ref{en}) is the kinetic energy at the distances $r <  \sqrt{1+k^2a^2}/k$, where the helical deformation increases the length of the vortex line by the factor $\sqrt{1+k^2a^2}$, while the second term is the kinetic energy at distances $r>\sqrt{1+k^2a^2}/k$, where helical deformation does not affect the  velocity $v=\kappa/2\pi r$ determined only by the circulation quantum. 

In general a momentum of a vortex line is given by \cite{Saf}
\be
\bm P =\frac{\rho_s\kappa }{ 2} \int [\bm R_L \times \upd \bm l]\, .
  \ee{mg}
Here $\upd  \bm l=\hat s \upd l$, and $\upd l$ is the element of the vortex line length. The expression leaves the vortex momentum undefined (similar to an undefined dipole moment of a single electron). 
But physically only the difference of the momenta of a distorted and a straight vortex along the chosen vertical axis is of importance. Then 
\be
P_z=\left.\frac{\rho_s \kappa L}{ 4\pi} \int s_\phi\sqrt{1+\left(\frac{\upd z}{ r\,\upd \phi}\right)^2}r^2 \, \upd \phi \right|_{r=a}= \rho_s \kappa L\frac {  a^2k}{ 2}\, .
  \ee{P}
This momentum is nothing else as the momentum of the constant vertical velocity field $\kappa k /2\pi$ inside the  cylinder of radius $a$ around which the vortex line is winding.  The winding vortex line with tilted circulation vector produces  a jump of the vertical velocity equal to  $\kappa k /2\pi$ in average, and the latter is equal to the vertical velocity inside the helix under natural assumption that there is no velocity outside the helix. Though the vertical velocity determines the whole momentum,  the kinetic energy of the vertical flow is ignored in the local induction approximation since it has not a large logarithmic factor. But we shall take the kinetic energy of the vertical flow into account studying the case when the strict local-induction approximation becomes invalid (see  below). Similarly, calculating the $z$ component  $M_z$ of the angular momentum of the velocity field induced by the helical vortex only the difference between the angular momenta of  a distorted and a straight vortex is of importance, and 
\be
M_z=-\frac{\rho_s \kappa L}{ 2}  a^2\, .
  \ee{PM}
Note a negative sign in this expression. Indeed, the angular momentum of a vortex line displaced from the container axis is less than the angular momentum of a vortex line exactly along the container axis. The latter is very large and depends on the container radius, but does not affect the outcome of the analysis determined only by momentum difference. 

\section{Helical-vortex ring: canonical formalism}

\begin{figure}
 \onefigure[scale=0.65] {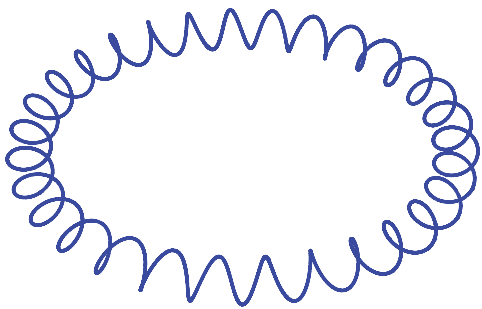}%
 \caption{Helical-vortex ring}  \label{f3}
 \end{figure}

Now let us roll up the helical vortex into a ring of the radius $R =L/2\pi$. The helix makes $n$ turns around the vortex ring (fig.~\ref{f3}). It is assumed that the radius $a$ of the helix is much smaller than $R$ but may be of the same order or larger than the pitch $2\pi R/n$ of the helix (the Kelvin wavelength). We shall derive the velocity of the ring in the absence of dissipation using the canonical relation for the ring velocity, which takes into account that the  coordinate along the ring axis and the linear momentum along the same axis constitute  a pair of  canonically conjugate Hamiltonian variables:
\be
v_r = \frac{\partial E }{ \partial P} = \frac{\partial E/\partial R }{ \partial P/\partial R }\, .
  \ee{ringV}
The momentum $P=\pi \rho_s \kappa R^2$ related with the ring is not affected by helical deformation  of the  vortex line. The expression (\ref{en}) for the energy given in the previous section can be used now after replacing $L$ by $2\pi R$ and $k$ by $n/R$. So the energy is 
\bem
E=\frac{\rho_s \kappa^2  }{ 2} \left(\sqrt{R^2+n^2a^2}\ln\frac{R}{ n r_c}
+R\ln\frac{nR _r}{ \sqrt{R^2+n^2a^2}}\right)\, .
  \eem{enR}

But now the system has two degrees of freedom, and one should decide what is going on with the vortex helix if the ring radius varies. Certainly the number $n$ of turns cannot vary being a topological invariant. Another invariant is the angular momentum along the ring axis. For large ring radius $R\gg a$ it can be estimated from the component $P_z$ of the straight helical vortex, \eq{P}:
\be
M_{ax}=\pi  \rho_s \kappa  R  a^2n\, . 
  \ee{6}
Variation of the energy with respect to $P$ must be performed at fixed $M_{ax}$. So the growth  of $R$ is accompanied by a decrease of the helix radius $a$: $da/dR= -a/2R$. Keeping the most important logarithmic terms the ring velocity is 
\be
v_r =\frac { \kappa }{ 4\pi }\frac{1-n^2a^2 /2R^2}{ \sqrt{R^2+n^2a^2}}\ln\frac{\sqrt{R^2 +n^2a^2}}{ n r_c}\, .
  \ee{canon}
So  at large number of helix turns $n > R/a$ the sign of the velocity changes, and the ring starts to move in the direction opposite to its momentum. This is a phenomenon of the anomalous vortex-ring velocity, which is  here directly explained by the canonical equations of motion and the angular momentum conservation law. Expansion of \eq{canon} in $na/R$ yields  the perturbation-theory result by Kiknadze and Mamaladze \cite{KM}. The critical value of $na/R=\sqrt{2}$ at which the vortex ring stops is larger than $na/R=1$ given by the perturbation theory and is in  reasonable agreement with the numerical calculations with the Bio-Savart law giving the critical value $na/R=1.7$ at large $n$ \cite{BHT}.

The angular velocity of the helical-vortex ring also follows from the canonical relation: 
\be
\Omega =\frac{ \partial E}{\partial M_{ax}}=\frac{\nu_s n }{ R\sqrt{R^2+n^2a^2}}\, .
     \ee{7}
Now taking the partial derivative the linear momentum $P$ ({\em i.e.}, the average radius $R$) must be fixed. Rotation with angular velocity $\Omega $ corresponds to translation of the straight helical vortex, so  $\Omega =v_z/R $, where $v_z$ is  given by \eq{disp} but with $k$ replaced with $n/R$.

\section{Helical-vortex ring:  general analytical solution}

The problem of a helical-vortex ring in the local induction approximation can be solved analytically without any assumption on the ratio $a/R $. The shape of the helical-vortex ring is determined in the cylindrical coordinates by the two functions $r(\phi,t)$ and $z(\phi,t)$. We look for an automodel solution $r(\phi,t)=r(\phi-\Omega t)$ and $z(\phi,t)=z(\phi-\Omega t)+v_r t$, which describes stationary translation and rotation of the ring with the linear velocity $v_r$ and the angular velocity $\Omega $ respectively.
The vector equation (\ref{EqCur}) of motion reduces to the two equations for the two time-independent  functions 
$r(\phi)$ and $z(\phi)$:
\bem
   \Omega r\frac {\upd r}{ \upd \phi} +\frac{\upd }{ \upd \phi}
 \left[ 
\frac {\nu_s  \upd z/\upd \phi}{ r\sqrt{1+(\upd r/r\, \upd \phi)^2+(\upd z/r\, \upd \phi)^2}} 
 \right] =0\, , 
  \eem{8}
\bem
- v_r r\frac {\upd r}{  \upd \phi} +\frac {\upd }{ \upd \phi}
 \left[ \frac {\nu_s r}{ [1+(\upd r/r\, \upd \phi)^2+(\upd z/r\, \upd \phi)^2]^{1/2}}  \right] =0\, .
  \eem{9}
These are two equations of  balance for forces along the axis $z$
and for moments around the axis $z$ (azimuthal forces) applied to an element of the vortex line: The left-hand sides present the Magnus forces, whereas the right-hand sides are the line-tension forces proportional to the components of the curvature vector in the cylindric coordinate frame. The first integration of the equations is straightforward:
\bem
\left(\frac{\upd r}{ \upd \phi}\right)^2 \nonumber \\
=\frac{r^4[4\nu_s^2-\Omega^2 (r^2-R^2)^2]}{ [v_r  (r^2-R_1^2)+R_1\sqrt{4\nu_s^2-\Omega^2 (R_1^2-R^2)^2}]^2}-r^2\, ,
\nonumber \\
\left(\frac{\upd z}{ \upd \phi}\right)^2=\frac{r^4\Omega^2 (r^2-R^2)^2}{  [v_r  (r^2-R_1^2)+R_1\sqrt{4\nu_s^2-\Omega^2 (R_1^2-R^2)^2}]^2}\, .
   \eem{eqR}
At integration  two  boundary conditions were used: (i) $dz/d\phi=0$ at $r=R$ [the maximum of the function $z(\phi)$],  and  (ii) $dr/d\phi =0$ at $r=R_1<R$   [the minimum  of the function $r(\phi)$]. 
The maximum  of the function $r(\phi)$ (the  the vortex line position most distant from the ring axis) is at the distance $r=R_2$ given by
\begin{center}
\textit{See equation \ref{long}  next page}
\end{center}
\begin{widetext}
\begin{equation}
R_2^2=R^2-\frac{1}{ 2}\left( R_1^2+\frac{v_r^2 }{ \Omega^2}\right)
+\sqrt{\frac{1}{ 4}\left( R_1^2+\frac{v_r^2 }{ \Omega^2}\right)^2 
+\left( R_1^2   -\frac{v_r^2 }{ \Omega^2}\right)(R^2-R_1^2)
+\frac{4\nu_s^2}{ \Omega^2}-\frac{4 v_r  R_1 \nu_s}{  \Omega^2}\sqrt{1-\frac{\Omega^2 (R_1^2-R^2)^2}{ 4\nu_s^2}} }\, .
\label{long}\end{equation}
\end{widetext}
The second integration yields expressions for  functions $\phi(r)$ and $z(r)$ via elliptic integrals:
\bem
\phi(r)=\frac {v_r(R_1^2+R_3^2)}{  \Omega R_3^2 \sqrt{R_2^2+R_3^2}}\left[F(\kappa, m)
\right. \nonumber \\  \left.
-\Pi \left( -\frac{R_3^2}{ R_1^2}m,\kappa,m\right)\right]
+\frac{\sqrt{4\nu_s^2-\Omega^2 (R_1^2-R^2)^2}}{ \Omega R_1 R_3^2 \sqrt{R_2^2+R_3^2} }
\nonumber \\  \times
 \left[(R_1^2+R_3^2)\Pi \left( -\frac{R_3^2}{ R_1^2}m,\kappa,m\right)-R_1^2F(\kappa, m) \right]\, ,
   \eem{phi}
    \bem
z(r)=\frac{(R_1^2+R_3^2) \Pi (m,\kappa,m)-(R^2+R_3^2 )F(\kappa,m)}{ \sqrt{R_2^2+R_3^2}}\, .
   \eem{10}
Here
\bem
 m=\frac{R_2^2-R_1^2}{ R_2^2+R_3^2}\,,~~\kappa=\arcsin\sqrt{\frac{r^2-R_1^2}{ m 
(r^2+R_3^2)}}\,,~~
R_3^3=R_1^2
 \nonumber \\
 +R_2^2-2R^2+\frac{v_r^2 }{ \Omega^2}\,,\qquad F(\phi,m)=\int_0^\phi\frac{\upd \alpha}{ \sqrt{1-m\sin^2 \alpha}},
   \nonumber \\
 \Pi(l,\phi,m)=\int_0^\phi\frac{\upd \alpha}{ (1-l\sin^2 \alpha)\sqrt{1-m\sin^2 \alpha}}\, .
     \eem{par}
Not all necessary boundary conditions were still satisfied. In order to obtain a continuous curve spiraling along toroidal surface one should require that at the maximum $r=R_2$ the coordinate $z$ returns to the same value as at the minimum $r=R_1$. The latter chosen to be zero this yields the condition $z(R_2)=0$.  The second condition is dictated by a chosen  period of the helix around the ring and is given by 
$\phi(R_2)=\pi/n$, where $n$ is the number of turns by the helical vortex along the ring. All together this provides a full analytic solution of our problem. 

\begin{figure}
 \onefigure[scale=0.4]{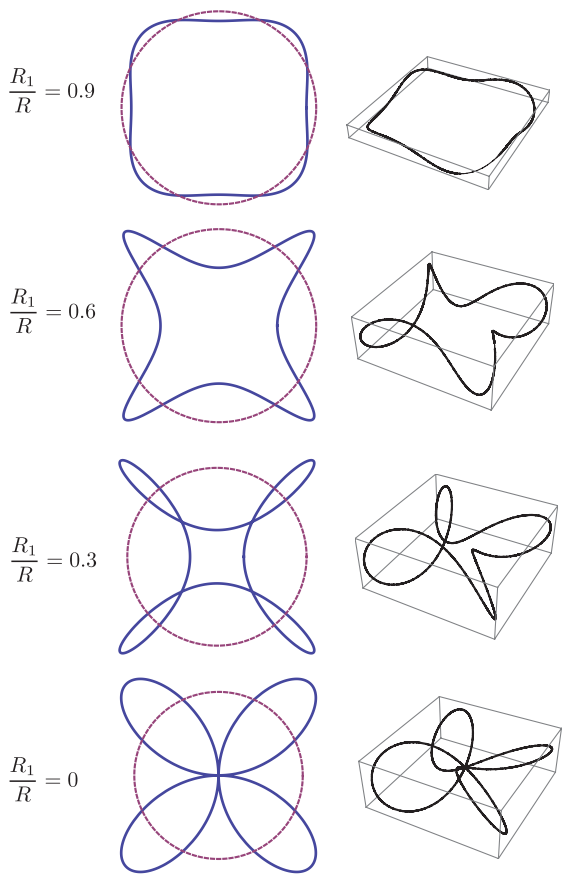}%
 \caption{Helical-vortex rings of four-fold symmetry $n=4$  for various ratios $R_1/R$. The left column shows projections of rings on the plane normal to the ring axis (the axis $z$). Dashed lines show unperturbed ideally circular rings of the radius $R$. The right column shows 3D images of rings.}  \label{f4}
 \end{figure}

\begin{figure}
 \onefigure[scale=0.3]{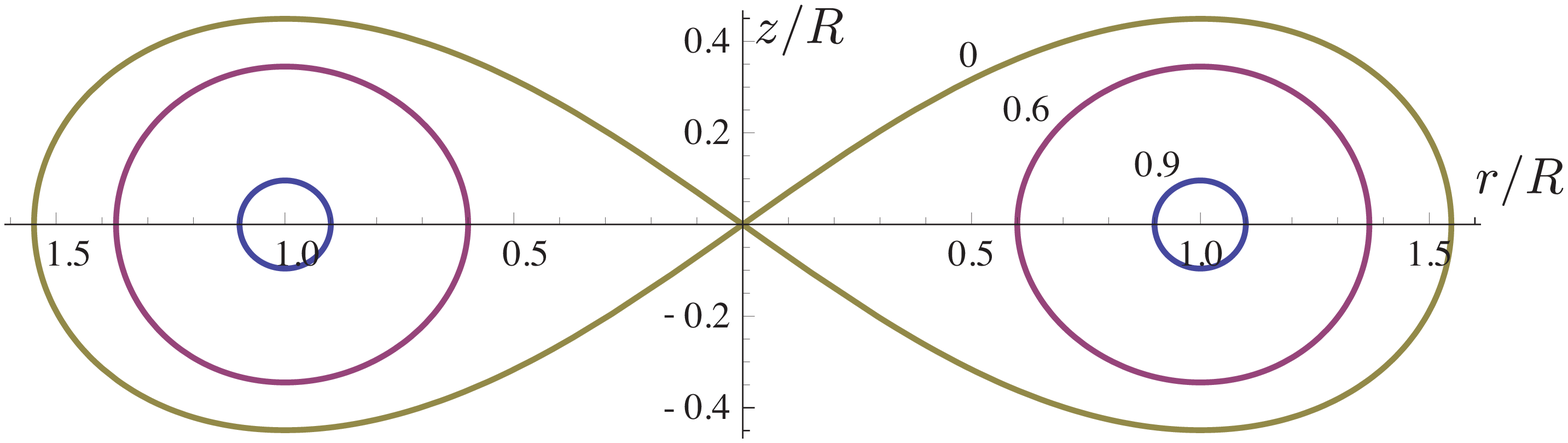}%
 \caption{Axial cross-section of toroids over which vortex lines wind for the case
 of four-fold symmetry $n=4$ and ratios $R_1/R=0.9$ ,0.6, and 0 (shown near the curves). The cross-section varies from two ideal circles at $R_1\to R$  to the figure eight at $R_1\to 0$. The latter corresponds to the bifurcation point where reconnection to a few separate loops may occur. }  \label{f5}
 \end{figure}

First one can check the limit of a weak Kelvin wave propagating around the ring (the linear theory). In this limit $a=R-R_1=R-R_2$ is very small. Then our solution yields that the translational velocity $v_r = \nu_s/R$ of the ring does not differ from that of an unperturbed ring. At the same time the angular velocity of the helix is given by $\Omega = \nu_s \sqrt{n^2-1}/R^2$, like that predicted by the linear theory of the Kelvin wave along the circular vortex loop. The most interesting case for us is  when the amplitude of the Kelvin wave is still small compared with the ring radius but not small compared with the Kelvin wavelength $2\pi R/n$. This requires large winding numbers $n\gg 1$. In this limit one obtains exactly the same result as given by \eq{canon} derived from the canonical formalism. In order to demonstrate this one should keep terms of the second order in $a=R-R_1$ and therefore expand the elliptic integrals in \eq{par} up to terms $m^2$. For not large wave numbers $n$ and large amplitudes $a=R-R_1$ one should use  values  of unexpanded elliptic integrals in calculations. Figure~\ref{f4} shows helical-vortex rings of wave number $n=4$  for various ratios $R_1/R$. The vortex line is winding around a surface of a toroid, which changes from an ideal torus at $R_1\to R$  with circular cross-section to a toroid obtained by revolution of a figure of eight at $R_1\to 0$ (fig.~\ref{f5}).

Our analysis demonstrates the existence of an automodel solution when the whole motion of the helical-vortex ring reduces to its pure translation and rotation without any variation of its shape.  Meanwhile, Barenghi {et al.} \cite{BHT} reported that in their numerical calculations using the Bio--Savart law they detected oscillations of the total length of the vortex line  with time. It is difficult to imagine that the automodel solution exists only in the local-induction approximation, especially keeping in mind  that calculations on the basis of  the Bio--Savart law confirm its existence for straight helical vortices \cite{Oku}. Apparently, in simulations by  Barenghi {et al.} the vortex line length oscillates because of excitation of more than one Kelvin mode.

\section{Beyond the local induction approximation}

All the results received above were obtained strictly within the local-induction approximation, which takes into account only the vortex velocity induced by the close segment of the vortex line and determined by its curvature.. Meanwhile, in a helical vortex of low pitch $2\pi /k \ll a$ [fig.~\ref{f2}(b)] distant parts of the same vortex line (distant in the sense of the distance measured along the vortex line) approach to a given point of the vortex line to distances much shorter than the curvature radius. Then the local-induction approximation in its strict meaning becomes invalid: one must take into account velocity induced by the  other turns of the spiral. The most rigorous way to deal with such cases is using the Bio--Savart law.  In classical hydrodynamics an analytical solution of the  Bio--Savart equations  for helical vortices was obtained in the form of  infinite series of Bessel functions (Kapteyn series)   \cite{Oku}. On the other hand, one may suggest a much simpler but still reasonably accurate approach, which takes into account  the  velocity induced by distant parts of the vortex line by   modification or  extension of the local-induction approximation,
 
We start from the straight vortex shown in fig.~\ref{f2}(b). At low pitch the surface of the cylinder around which the vortex line is winding can be considered as a continuous vortex sheet with the jump of the vertical velocity component $\kappa k/2\pi$. Assuming  that there is no velocity outside the helix, but only inside, one obtains the kinetic energy of axial (vertical) motion per unit length: $\rho_s \kappa^2 k^2 a^2/8\pi$. Adding this energy to the energy  in the local-induction approximation given by \eq{en} one obtains
\bem 
E=\frac{\rho_s \kappa^2 L}{ 4\pi} \left(\sqrt{1+k^2a^2}\ln\frac{1}{ kr_c}+\ln\frac{kR_c }{ \sqrt{1+k^2a^2}}
\right. \nonumber \\ \left.
+\frac{k^2a^2}{2}\right)\, .
  \eem{enC}
The added kinetic energy (the last term in the parentheses) is related to the velocity induced by all turns of the vortex line around the cylinder of radius $a$. It is beyond the strict local-induction approximation since it does  not  contain a large-logarithm factor. But it provides the most essential contribution at $ak \gg |\ln(kr_c)|$.  In this limit this approach must yield the same results as the Bio--Savart law. So the expression (\ref{enC}) is a reliable interpolation between two limits $ak \ll |\ln(kr_c)|$ (local-induction limit) and $ak\gg |\ln(kr_c)|$ (continuous-vortex-sheet limit). The extended local-induction approximation does not change expressions for the linear momentum $P_z$ [\eq{P}]  and the angular momentum $M_z$ [\eq{PM}] since at their derivation the  local-induction approximation    was not used.

Using the expression (\ref{enC}) for the energy  in  the canonic expressions for the translational and angular velocities [\eq{can}] one obtains:
\be 
v_z=\frac{\omega}{ k} =\frac {\nu_s k }{ \sqrt{1+k^2a^2}}+\frac{\kappa k}{4\pi}\,,
  \ee{dispC}
\be
\Omega_z  =-\frac {\nu_s k^2 }{ \sqrt{1+k^2a^2}}-\frac{\kappa k^2}{4\pi}\, .
  \ee{rotC}
Note that the continuous-vortex-sheet contribution $\kappa k/4\pi$ to the translational velocity is an average between the zero velocity outside and the velocity  $\kappa k/2\pi$ inside the cylindric vortex sheet like in the case of plane vortex sheets well known in hydrodynamics.

A similar modification of the local-induction approximation is possible for helical-vortex rings. The added continuous velocity  is now an azimuthal velocity inside the helix formed along the ring. This leads to modification of the expression \eq{canon} for the ring translational velocity:
\be
v_r =\frac { \kappa }{ 4\pi }\frac{1-n^2a^2 /2R^2}{ \sqrt{R^2+n^2a^2}}\ln\frac{\sqrt{R^2 +n^2a^2}}{ n r_c}-\frac{\kappa n^2a^2}{8\pi R^3}\, .
  \ee{canonC}
A negative sign of the new non-logarithmic term demonstrates that the reverse motion of the vortex ring due to its helical deformation remains also in the continuous-vortex-sheet limit.

The principle, at which  the extended local-induction approximation is based, is not new. The well-known Hall--Vinen--Bekarevich--Khalatnikov theory for rotating superfluids \cite{D,RMP} was based on the same principle: The  logarithmically large self-induction line-tension contribution to the vortex-line velocity is combined with the velocity induced by other vortices in the vortex array, which is approximated by continuous vorticity. In our case there is no other vortices, but the continuous vorticity is introduced as an approximation for the effect of other turns of the same vortex line.

\section{Other extensions and implications of the analysis}

Studying the helical-vortex rings the analysis addressed the case of a single Kelvin mode with fixed wave number. A natural extension of the analysis is the case of the ensemble of Kelvin modes assuming that the effect of various modes is additive \cite{Bra}, {\em i.e.}, replacing $n^2a^2$ by $\sum\limits _n n^2 a_n^2$. Then ${\cal L}=2\pi \sqrt{R^2+\sum\limits _n n^2 a_n^2}$ is the total length of the vortex line increased by the vortex line fluctuations. 

We have obtained that elongation of the vortex line decreases the translational ring velocity, but this conclusion is not universal being related with conservation of the angular momentum at propagation of an isolated single vortex ring. 
Dynamics of vortex rings is widely exploited in the theory of superfluid turbulence.
One may not expect that in a vortex tangle with many interacting rings the conservation law for a particular vortex  is applicable. On the other hand, a possibility can be considered that interaction of Kelvin modes on various rings establishes some fixed average energy   per unit length of any  vortex line. Then  the partial derivative of the energy with respect to the ring radius must be calculated at fixed ratio $\sum\limits _n n^2 a_n^2/R$ and the translational velocity $v_r = \nu_s {\cal L} /2 \pi R^2$ exceeds that of the ideal circular vortex ring. One may consider this as renormalization of the line tension parameter $\nu_s$ by the factor  ${\cal L} /2\pi R$ taking into account elongation of the vortex line by short-wavelength Kelvin modes.

\section{Conclusions}
The Letter studies dynamics of helical vortices and helical-vortex rings in the local-induction approximation either analytically solving directly the equations of motions of a vortex line or using canonical relations following from the Hamiltonian equations of motion. The analytical solution of equations of motions fully confirms the results following from simple canonical relations. The analysis based on these relation explains the origin  of the anomalous  vortex-ring velocity, which does not deserves an adjective ``anomalous'' anymore.  Suppression of the ring velocity and  eventual inversion of its direction are related with conserved angular momentum around the ring axis, which is generated by the helical deformation of the vortex ring.  The analysis goes beyond the local-induction approximation and suggests the extended local-induction approximation, which agrees with the exact solution for  helical vortices and helical-vortex rings in the limit when the  vortex line winding around a cylinder (helical vortex) or a torus (helical-vortex ring) reduces to a sheet with uniformly distributed vorticity.

\acknowledgments
The work was supported by the grant of the Israel Academy of Sciences and Humanities.

\end{document}